\documentstyle[aps,prl,multicol,epsf]{revtex}
\renewcommand{\narrowtext}{\begin{multicols}{2} \global\columnwidth20.5pc}
\renewcommand{\widetext}{\end{multicols} \global\columnwidth42.5pc}
\renewcommand{\v}[1]{{\bf #1}}

\newcommand{\ba}{\begin{eqnarray}}
\newcommand{\ea}{\end{eqnarray}}
\newcommand{\be}{\begin{equation}}
\newcommand{\ee}{\end{equation}}

\begin{document}
\draft
\title{
Low-Frequency Optical Conductivity in \\ Inhomogeneous d-wave
Superconductors}
\author{Jung Hoon Han}
\address{Department of Physics, Konkuk University \\
1 Hwayang-Dong Kwangjin-Gu\\
Seoul, 143-701 Korea}

\maketitle \draft
\begin{abstract}
Motivated by the recent optical conductivity experiments on
Bi$_2$Sr$_2$CaCu$_2$O$_{8+\delta}$ films, we examine the possible
origin of low-frequency dissipation in the superconducting state.
In the presence of spatial inhomogeneity of the local phase
stiffness $\rho_s$, it is shown that some spectral weight is
removed from $\omega=0$ to finite frequencies and contribute to
dissipation. A case where both $\rho_s$ and the local normal fluid
density are inhomogeneous is also considered. We find an enhanced
dissipation at low frequency if the two variations are
anti-correlated.
\end{abstract}

\narrowtext

Identifying the nature of quasiparticles and their dynamics is an
important part of the overall understanding of the properties of
matter. In $d$-wave cuprate superconductors, gap-vanishing nodes
are responsible for an appreciable density of quasiparticles even
for a superconducting state. Optical conductivity measurements
carried out on YBCO crystal\cite{hosseini} shows that the
quasiparticle dynamics in the superconducting state obeys the Drude
form,
\begin{equation}
\sigma_{n}(\omega,T) = {{\rho_n (T) \tau_{n} (T)}\over
1+\omega^2\tau_{n}^2(T)},
\end{equation}
where $\rho_n (T)$ is the $T$-linear normal fluid fraction, and
$\tau_n (T)$ is the quasiparticle lifetime at a given temperature.
As the temperature is lowered $\sigma_n (\omega,T)$ is seen to
decrease due to vanishing quasiparticle density.

For Bi$_2$Sr$_2$CaCu$_2$O$_{8+\delta}$ (Bi:2212), on the other
hand, a similar microwave measurement shows substantial amount of
residual conductivity as $\omega\!\rightarrow\! 0$ and
$T\!\rightarrow\! 0$\cite{cambridge}, even though the operating
temperature and frequency ranges of the measurement were comparable
to those in Ref. \cite{hosseini}. It was recently shown that a good
fit to the optical conductivity data for slightly underdoped
Bi:2212 thin film (T$_c$=85K) can be achieved if one introduces
another term, denoted $\sigma_c$, in addition to the two-fluid
contribution in the conductivity\cite{corson}:
\begin{equation}
\sigma_c (\omega, T)= \kappa{{\rho_s (T)\tau_{c}}\over
1+\omega^2\tau_c^2}. \label{sigmac}
\end{equation}
The low-frequency conductivity $\sigma_c$ now {\it grows} as the
superfluid density $\rho_s (T)$ upon lowering the temperature.
Further studies showed that $\kappa$, which is a measure of the
amount of spectral weight removed from the condensate, grows
monotonically with increasing doping concentration.

Such behavior of the optical conductivity is unusual from the
simple two-fluid point of view, and deserves theoretical
attention\cite{barabasi,orenstein}. It is also essential to clarify
the source of discrepancy of the optical data among different
materials. In the context of Josephson-coupled superconducting
grains, Barabasi and Stroud showed that an inhomogeneous
distribution of the Josephson coupling strength $J$ in
superconducting arrays gives rise to a shift of spectral weight
under $\omega=0$ (the condensate fraction) to finite frequencies
with an amount proportional to $\langle (\delta
J)^2\rangle/\overline{J}$, where $\overline{J}$ is the mean phase
stiffness of the system\cite{barabasi}. Roughly speaking, the
inhomogeneity in $J$ gives rise to scattering of superfluid density
fluctuations (spin waves) which tends to move some of the states
from the condensate into the dissipative part.

Evidence in support of inhomogeneous nature of the superconducting
state in Bi:2212 has been mounting from STM images of the BSCCO
surface\cite{davis,kapitulnik,cren}. Theoretical consideration
shows that the randomness of the dopant oxygen, inherent in the
doping process for Bismuth compounds, does produce an
inhomogeneous hole distribution in the copper-oxygen
plane\cite{theory}. The zero-temperature phase stiffness is in
turn determined by the local hole density in cuprate
superconductors, and is very likely to be disordered. The purpose
of this paper is to calculate the optical conductivity induced by
the randomness in the local phase stiffnes $\rho_s ({\bf r})$.
Inherent in our model is the assumption that the inhomogeneity is
in fact present in the bulk, although the direct evidence from STM
is confined to the surface layer of the compound. Other evidence
in support of the bulk-disordered nature of the Bi:2212
superconductors can be found, for example, in transport, neutron
scattering, and ARPES measurements of Bi:2212 against a similar
measurement for YBCO\cite{davis}.

Our approach is based on the effective action for the phase of the
condensate order parameter, $\phi$. Starting from a Lagrangian
which gives two-fluid description of the low-energy excitations in
charged superfluids, we arrive at an effective action for $\phi$
alone through integrating out other degrees of freedom. A general
formula for the optical conductivity in the presence of phase
stiffness disorder is derived via the replica method, and using the
effective action for $\phi$, computed to lowest order in the
disorder strength.

\widetext Effective action of the phase-fluctuating
(2+1)-dimensional superfluid can be written down in the following
general form

\begin{eqnarray}
S&=&{1\over2} \int d^3 x d^3 y [K(x-y) \!+\!\delta \rho_s ({\bf r})
\delta(x\!-\!y)] ({\bf grad}
\phi\!+\!{\bf A})\!(x) ({\bf grad} \phi\!+\!{\bf A})\!(y) \nonumber \\
&+& {1\over2} \int d^3 x d^3 y (\partial_0 \phi \!+\!A_0)(x)
U^{-1}(x-y) (\partial_0 \phi \!+\! A_0 )(y).
\label{action}
\end{eqnarray}
\narrowtext \noindent We denote the space-time coordinates as $x$
and $y$. All the linear-response properties are summarized by the
non-local kernels $K(x-y)$ and $U^{-1}(x-y)$, which will be
specified later. As the above action shows, the stiffness disorder
$\delta\rho_s$ is quenched, and does not fluctuate in time. We
further assume that $\delta \rho_s ({\bf r})$ obeys the Gaussian
distribution

\begin{equation}
P[\delta\rho_s ({\bf r})]\propto \exp\left(-{1\over 2g}\int d^2 \v
r [\delta \rho_s ( \v r)]^2\right).
\end{equation}
\noindent The disorder average of the partition function $Z$ can be
carried out in the usual manner by first re-writing $\ln Z$ as
$(Z^N -1)/N$, then integrating over $\delta\rho_s (\v r)$. This
leads to a replica action which is $N$ copies of the action in Eq.
(\ref{action}) without $\delta\rho_s$, plus the disorder-induced
term
\begin{eqnarray}
&&-{g\over2}\int d^2{\bf r} d\tau d\tau'
\sum_{m=1}^N ({\bf grad} \phi_m \!+\!{\bf A} )^2
({\bf r}\tau) \nonumber \\
&&~~~~~~~~\times\!\sum_{n=1}^N ({\bf grad} \phi_n \!+\!{\bf A})^2
({\bf r}\tau'),
\label{disorder-in-action}
\end{eqnarray}
where $m,n$ are the replica indices.

The current-current
correlation function $\Pi_{ij}$ follows from
\begin{equation} \Pi_{ij}
({\bf r}_1 \!-\!{\bf r}_2, \tau_1 \!-\!\tau_2)=-\lim_{N\rightarrow
0} {\partial^2 \langle \ln Z \rangle\over {\partial A_i ({\bf
r}_1\tau_1)
\partial A_j ({\bf r}_2\tau_2)}},
\label{Pi_ij}
\end{equation}
using the replica action defined above. Also in an optical
experiment we are interested in the transverse piece of $\Pi_{ij}$.
Contributions of Eq. (\ref{disorder-in-action}) to $\Pi_{ij}$ is
two-fold. First, expanding Eq. (\ref{disorder-in-action}) one
obtains (dropping replica indices) $\sim ({\bf grad}\phi)^2 ({\bf
grad}\phi)^2$ term which, to first order in $g$, can be replaced by
$\sim \langle ({\bf grad}\phi)^2 \rangle ({\bf grad}\phi)^2$ and
modifies the kernel $K(k)$. Secondly there is $\sim (\v A \cdot
{\bf grad}\phi)^2$, which will contribute $\sim \langle {\bf
grad}\phi\,\, {\bf grad}\phi\rangle$ in the current-current
correlation function. As a result the following form is obtained
for the transverse part: \be \Pi_t (k) = K(k)-4g G(i\omega_m ).
\label{Pi_t}\ee Here, $K(k)$, $k=({\bf k},i\omega_m )$ is the
Fourier transform of the kernel $K(x-y)$, and
\begin{equation}
G(i\omega_m) = \int {d^2 {\bf k}\over (2\pi)^2} {\bf k}^2 \langle
\phi(-k)\phi(k)\rangle. \label{G_w}
\end{equation}
The momentum integral is restricted to $|{\bf k}|<\Lambda$, which
is the inverse of the typical length scale of the inhomogeneity.
The boson-boson correlator is given by \be \langle
\phi(-k)\phi(k)\rangle^{-1}\equiv s(\v k,i\omega_m )=\v k^2
K(k)+\omega_m^2 /U(k).\label{boson-boson}\ee The optical
conductivity is in turn obtained from $\Pi_t$ through \be\sigma
(\omega)= \left.{\Pi_{t}({\bf k}=0,i\omega_m )\over\omega_m} \right
|_{i\omega_m \!\rightarrow\! \omega+i\delta}. \label{sigma} \ee

Equations (\ref{Pi_t})-(\ref{sigma}) form the basis for calculating
the optical conductivity of a given superfluid in the presence of
phase stiffness disorder. The properties of a superfluid are
encoded in the linear response functions $K(k)$ and $U^{-1}(k)$,
which in turn can be derived from a specific microscopic model.
Once they are determined, optical conductivity immediately follows
from the above formulae.

We model the low-energy dynamics of a superconductor as charged
two-fluid system described by the action

\begin{eqnarray} &&S={1\over 2K}{\bf J}_s^2 +
{u\over 2} (\rho_s \!-\!\overline{\rho})^2 +
i(J_{s\mu}+J_{n\mu})(\partial_{\mu}
\phi +A_{\mu}) \nonumber \\
&&~~~~~~~~+ S_n + {\epsilon_0 \over {8\pi}} ({\bf grad}A_0 )^2.
\label{charged-action}
\end{eqnarray}
In the case of $d$-wave superconductors, the quasiparticle action
$S_n$ is that of Dirac particles. The super- and normal-fluid
three-currents are respectively denoted as $J_{s\mu}=(\rho_s
\!-\!\overline{\rho}, \v J_s)$, and $J_{n\mu}$ ($\mu=0,1,2$). One
recovers the current conservation law, $\partial_\mu
(J_{s\mu}+J_{n\mu})=0$, upon integrating out the phase $\phi$.
There is no separate conservation of $J_{s\mu}$ or $J_{n\mu}$ alone
since conversion of superfluid into normal fluid and vice versa can
take place through scattering processes. The gauge field action
$\epsilon_0 ({\bf grad} A_0 )^2/8\pi$ generates the long-range
Coulomb interaction between charges. The bare stiffness and
compressibility of the superfluid are given by $K$ and $u^{-1}$,
respectively.

On integrating out the normal current $J_{n\mu}$ and the
supercurrent $J_{s\mu}$ successively we arrive at an effective
action:

\begin{eqnarray} S&=&{K\over2}({\bf grad} \phi \!+\! {\bf
A})^2 + {1\over
2u}(\partial_0 \phi \!+\! A_0 )^2 \nonumber \\
&+& {1\over 2} (\partial_{\mu} \phi + A_\mu )
\kappa_{\mu\nu}(\partial_{\nu} \phi + A_{\nu}) +{\epsilon_0 \over
{8\pi}} ({\bf grad} A_0 )^2.
\label{tfm}
\end{eqnarray}
The result of integration over the normal current $J_{n\mu}$ has
been expanded up to quadratic order in $\partial_\mu\phi+A_\mu$,
therefore $\kappa_{\mu\nu}$ is the correlation function of the
normal 3-current, $\langle J_{n\mu}J_{n\nu}\rangle$.  We further
write $\kappa_{00}=\chi_n$, and
$\kappa_{11}=\kappa_{22}=\kappa_n$. For the off-diagonal terms,
$\chi_{0i}=0$ due to the absence of time-reversal-symmetry
breaking, and $\chi_{i\neq j}=0$ assuming a typical nonmagnetic
impurity scattering of quasiparticles.

Let us first review the case of homogeneous superconductor. We set
$K=\rho_s$, and write $\kappa_n$ in terms of the quasiparticle
conductivity $\sigma_n$ as
\begin{equation}
\kappa_n =-\rho_n (T)+|\omega_m |\sigma_n. \label{sigma_n}
\end{equation}
With this substitution, and in the absence of disorder, the
transverse current-current correlation function $\Pi_t$ becomes
$\rho_s -\rho_n(T)+|\omega_m |\sigma_n$. Once $\rho_s -\rho_n (T)$
is identified as the finite-temperature superfluid density $\rho_s
(T)$, the conductivity at long wavelength (${\bf k}=0$) is obtained
from Eq. (\ref{sigma})
\begin{equation}
\sigma(\omega, T) ={{i\rho_s (T)} \over {\omega+i\delta}} +
\sigma_n. \label{tf}
\end{equation}
This is nothing but the two-fluid result for the conductivity. We
have thus demonstrated that our starting point in Eq.
(\ref{charged-action}) is indeed consistent with the two-fluid
picture.

Finally upon integrating out the scalar potential $A_0$ in action
(\ref{tfm}), we obtain the boson propagator

\begin{eqnarray} s(\v k,i\omega_m)
&=&(\rho_s (T) +|\omega_m |\sigma_n ) {\bf
k}^2 \nonumber \\
&&~~~~~~+\omega_m^2 { {u^{-1}+\chi_n } \over {1+4\pi(u^{-1}+\chi_n
)/\epsilon_0 \bf{k}^2}}. \label{boson_propagator}
\end{eqnarray}
Knowledge of this quantity is not necessary for the evaluation of
the conductivity in the homogeneous superconductor, but will be
important for the disorder-induced contribution as one can see from
Eqs. (\ref{Pi_t})-(\ref{boson-boson}). One can in fact identify
$\rho_s (T) + |\omega_m |\sigma_n$ and $u^{-1}+\chi_n$ as
current-current and density-density correlation functions of the
combined superfluid + normal system. Inclusion of the Coulomb
interaction renormalizes the latter quantity in the RPA
manner\cite{randeria}. We can now proceed using this boson
propagator to understand the effects of inhomogeneity. To clarify
the physical content of the expression, Eq.
(\ref{boson_propagator}), we introduce
\begin{equation} \gamma=\gamma_1
+i\gamma_2
=\frac{4\pi}{\epsilon_0}+\frac{ {\bf
k}^2}{u^{-1}+\chi_n},
\end{equation}
and write (after $i\omega_m\rightarrow \omega+i\delta)$ \be s(\v
k,\omega) = {{\bf k}^2 \over \gamma} \left[\gamma(\rho_s
(T)\!-\!i\omega\sigma_n) \!-\! (\omega\!+\!i\delta)^2\right]. \ee
Without $\sigma_n$, the terms inside the bracket gives the
Josephson plasma dispersion of the charged superfluid. With the
quasiparticles present, damping occurs in the plasma mode with a
characteristic lifetime $\sim\sigma_n/\rho_s (T)$.

When $s(\v k,\omega)$ is substituted in Eq. (\ref{Pi_t}) the
disorder-induced conductivity becomes \ba \sigma_1
(\omega)&=&-{{g\Lambda^2} \over
{\rho_s(T)}}\delta(\omega)+\sigma'_1(\omega) \nonumber \\
\sigma'_1 (\omega)&=&-\frac{4g}{\omega}{\rm Im}\left[\int\frac{d^2
{\bf k}}{(2\pi)^2}\frac{\gamma }{\omega^2\!-\!\gamma(
\rho_s(T)\!\!-\!\!i\omega \sigma_n ) }\right]. \label{here} \ea For
convenience we have written down only the real part. The amount of
spectral weight shift away from the condensate is indeed
proportional to
$\langle\delta\rho_s^2\rangle/\rho_s(T)$\cite{barabasi}, since
$g\Lambda^2\sim \langle\delta\rho_s^2\rangle$. They will re-appear
at non-zero frequencies in $\sigma'_1 (\omega)$. We examine the
behavior of $\sigma'_1 (\omega)$ under a number of circumstances.
\\

I) Zero quasiparticle dissipation: Take $\sigma_n = 0$. In this
case,

\begin{equation} \sigma'_1 (\omega)\approx
-4g\omega \int {{d^2 {\bf k}} \over{(2\pi)^2}} {\gamma_2
\over{[\omega^2 \!-\! \gamma_1 \rho_s (T) ]^2 + [\gamma_2 \rho_s
(T)]^2}}. \label{disorder-induced-sigma}
\end{equation}
For small $\gamma_2$, the integrand is narrowly centered around the
plasma frequency, $\omega^2\approx \gamma_1 \rho_s$, and zero
otherwise. Hence, the dissipation arises only if $\omega$ exceeds a
characteristic value comparable to Josephson plasma frequency. The
experimental observation of low-frequency dissipation\cite{corson}
is therefore impossible to explain in this case.
\\

II) Finite quasiparticle dissipation: We are after all interested
in frequency ranges much smaller than the plasma frequency. In
this case, the integrand appearing in Eq. (\ref{here}) is
simplified to $(i\omega\sigma_n-\rho_s (T))^{-1}$ and

\be \sigma'_1 (\omega) \approx{{g\Lambda^2} \over \pi} {\sigma_n
\over{\rho_s (T)^2 +\sigma_n^2 \omega^2}}.\label{sigmac2}\ee This
is a Lorentzian centered at $\omega=0$, with the width governed by
$\rho_s (T)/\sigma_n$.  By integrating this expression over
positive $\omega$, one finds that all of the spectral weight
removed from the condensate indeed re-appears in $\sigma'_1
(\omega)$.

We have thus demonstrated the existence of a new low-energy
dissipation term, Eq. (\ref{sigmac2}), for the superfluid which was
absent in the conventional two-fluid picture. Presence of this
effect rests on 1) residual quasiparticle dissipation
($\sigma_n\neq 0$) and 2) inhomogeneity in the superfluid phase
stiffness ($g\Lambda^2
>0$). Given that the doping process almost always results in an
inhomogeneous distribution of holes, we expect our result to hold
for a variety of other cuprate superconductors.

Up to this point it is assumed that $\sigma_n$ remains homogeneous
throughout the system, while the local superfluid density undergoes
a spatial fluctuation. It is certainly possible, however, that the
local normal fluid density is disordered as well as $\rho_s$. For a
sufficiently smooth variation, we can treat the local value of the
quasiparticle conductivity $\sigma_n (\v r)$ to be proportional to
$\rho_n (\v r)$. In particular we will consider a situation where
the variation of normal fluid density is tied to that of the
superfluid by $\delta\rho_n (\v r)=\lambda \delta\rho_s (\v r)$,
with $\lambda$ taking on positive (correlated) or negative
(anti-correlated) values. The quasiparticle conductivity is
accordingly \be \sigma_n (\v r)=\sigma_n +\lambda\delta\rho_s (\v
r)\tau_n.\ee We assume that the quasiparticle conductivity remains
in the d.c. limit, $\omega\tau_n \ll 1$.

Replica averaging over $\delta\rho_s (\v r)$ can be carried out
straightforwardly as before and yields the following complex
conductivity due to inhomogeneity: \be \sigma (\omega)={-i\over
\omega\!+\!i\delta}{g\Lambda^2 \over \pi} {(1-i\lambda\tau_n
\omega)^2 \over \rho_s (T)-i\sigma_n \omega}.\ee Our previous
discussion belongs to $\lambda=0$. The real part is now given by
($\omega>0$) \be \sigma_{1}(\omega)\approx {g\Lambda^2 \over
\pi}{{\sigma_n - 2\lambda\rho_s (T)\tau_n}\over{\rho_s(T)^2
+\sigma_n^2 \omega^2}}, \label{sigmac3}\ee ignoring a $(\omega
\tau_n)^2$ correction. Compared to our previous result in Eq.
(\ref{sigmac2}), one finds an {\em increase} in the low-frequency
conductivity in the case of {\em anti-correlated} spatial
variations of the normal and superfluid densities ($\lambda<0$),
and a decreased conductivity otherwise. A similar conclusion has
been reached in a recent preprint by Orenstein\cite{orenstein}.

We now discuss how the formulas derived above, in particular Eqs.
(\ref{sigmac2}) and (\ref{sigmac3}), compare with the experiment.
The damping rate $\tau_c^{-1}\equiv \rho_s /\sigma_n$ turns out to
be $100$K $\approx 2$THz, based on the zero-temperature estimate
of $\rho_s$ and $\sigma_n$ from the microwave and terahertz
data\cite{corson}. This value is consistent with the measured
value of $1.5$THz\cite{corson}. If we re-cast Eq. (\ref{sigmac2})
in the form of Eq. (\ref{sigmac}) using the definition $\tau_c =
\sigma_n /\rho_s (T)$, we arrive at $\kappa= g\Lambda^2 /\pi\rho_s
(T)^2 \sim \langle \delta\rho_s^2 \rangle/\rho_s(T)^2$. The
significance of $\kappa$ is thus clarified, as the relative
fraction of the inhomogeneity present in the superfluid density.
Furthermore, re-writing Eq. (\ref{sigmac3}) in the same form gives
\be \kappa(\lambda)=\kappa(\lambda=0)\times\left(1-{2\lambda\rho_s
(T)\over \rho_n}\right).\label{kappa_lambda}\ee The
$T\rightarrow0$ estimate of the superfluid and normal fluid
densities turn out to be comparable in available
samples\cite{orenstein}, and the degree of correlation $\lambda$
between normal fluid and superfluid densities turn out to play an
important role in estimating $\kappa$. Experiments carried out on
Bi:2212 films of varying doping concentrations ranging from
underdoped (T$_c$=51K) to overdoped (T$_c$=75K) indicate that
$\kappa$ monotonically increases from 0.08 to 0.46\cite{corson}
with doping. Equation (\ref{kappa_lambda}) suggests that one way
to understand the observed behavior of $\kappa$ is to say that the
superfluid and normal fluid densities in the ground state become
more and more anti-correlated upon over-doping.

In summary, we have considered the optical conductivity response
of a charged superfluid in the presence of inhomogeneous phase
stiffness distribution. In the homogeneous system only the
quasiparticle excitations contribute to low-frequency dissipation.
On the other hand, we find that some spectral weight is removed
from the condensate and shows up as low-frequency dissipation due
to spatial fluctuations of the local phase stiffness. In the model
we considered, the low-frequency distribution follows a
Lorentzian. The spectral weight lost at $\omega=0$ is proportional
to $\langle \delta\rho_s^2 \rangle/\rho_s (T)$. In the case where
the normal fluid density $\rho_n$ is also disordered, the
low-frequency dissipation is still present, and is increased
relative to the $\delta\rho_n =0$ case provided $\delta\rho_n (\v
r) \propto -\delta\rho_s (\v r)$, i.e. they are anti-correlated. A
decreased conductivity is found if they are spatially correlated.

I wish to thank Dung-Hai Lee for suggesting this problem in the
first place, as well as for extensive discussion on the subject.
Many fruitful discussion with Seamus Davis, Joe Orenstein, and
Qiang-Hua Wang are gratefully acknowledged. This work is supported
by the Faculty Fund of Konkuk University for 2001.

\widetext
\end{document}